
\input phyzzx

\def\rho{\varrho}

\tolerance=500000
\overfullrule=0pt

\pubnum={\vbox to 1.8cm{\hbox{CPTH-A182.0892}\hbox{CERN-TH.6605/92}}}
\date={ August 1992}
\pubtype={}
\titlepage

\title{PAIR CREATION OF OPEN STRINGS IN AN ELECTRIC FIELD} \vskip
2.0cm \author{ C.Bachas }
 \address{ Centre de
Physique Th\'eorique, Ecole Polytechnique\break 91128 Palaiseau,
France} \vskip 1.0cm \centerline{\it and} \vskip 0.3cm
\author{ M. Porrati \footnote\dag{On leave of absence from INFN,
Sezione di Pisa,
Italy} \footnote*{Address after September 1: Department of Physics, New
York
University, 4 Washington pl., New York, N.Y. 10003\hfil\break \ }}
\address{ Theory Division, CERN \break CH-1211, Geneva 23, Switzerland}
\vskip 1.5cm
\abstract{ We calculate exactly the rate of pair production of open bosonic
and
supersymmetric strings
 in a constant  electric field. The rate agrees with Schwinger's classic
result in the
weak-field limit, but diverges when the electric field approaches  some
critical value of the  order of the string tension.
 } \endpage

\pagenumber=1

\noindent One of the most beautiful calculations in field theory is
Schwinger's
calculation of  the probability of pair creation in a constant electric
field
\REF\Schwinger{ J.Schwinger, Phys.Rev. {\bf 82} (1951) 664.} [\Schwinger].
The probability per unit volume and unit time reads
$$ w =  {2J+1\over 8\pi^3} \sum_{k=1}^{\infty} \ (-)^{(2J+1)(k+1)}\
\Bigl({eE\over k}\Bigr)^2 \exp\Bigl(-{k\pi M^2\over
\vert eE\vert}\Bigr), \eqno(1) $$
where  $e$, $J$ and $M$ are the charge, spin and mass of
the produced particles, and $E$ is the external field.
This is one of the few non-perturbative results which can be calculated
exactly in
field theory. The purpose of this paper is to compute this probability when
point
particles are replaced by open strings.

The motion of an
open string in a constant electromagnetic background can be solved exactly
because
the corresponding world-sheet theory is free \REF\Fradkin{E.S.Fradkin and
A.A.Tseytlin, Phys.Lett. {\bf 163B} (1985) 123.} \REF\Abel{A.Abouelsaood,
C.G.Callan,
C.R.Nappi and S.A.Yost, Nucl.Phys.{\bf B280} [FS18] (1987) 599.} [\Fradkin,
\Abel].
To be sure the electromagnetic field
distorts in principle the background metric, but this effect is of
higher order in the genus expansion and can
for our purposes be ignored if the string coupling is
vanishingly small.  The issue of pair
production in this context has been considered before by Burgess
\REF\Bur{C.P.Burgess, Nucl.Phys. {\bf B294} (1987) 427.} [\Bur]
who calculated the rate for bosonic strings, and in
the limit of small field only. Here we will give the exact
expression for arbitrary field strength,
and for both bosonic and supersymmetric strings. We will find in particular
that
 this rate diverges at the same limiting value of the field, at which the
 string develops a
classical instability and the Born-Infeld action becomes ill-defined
\REF\Russ{V.V.Nesterenko,
Int.J.Mod.Phys. {\bf A4} (1989) 2627.}
[\Bur, \Russ]. We will also give an integral representation for the induced
one-loop (``Euler-Heisenberg") Lagrangian, a piece of which has been
calculated
before in refs.
[\Fradkin, \Abel]. \vskip 0.3cm

The action of the open bosonic string in a constant electromagnetic
background
and in conformal gauge reads
$$\eqalign{
S_{bos} = -{1\over 4\pi\alpha'} \int d\sigma & d\tau \  \partial_a X^{\mu}
\partial^a X_{\mu} \cr & + {1\over 2} e_1 \int d\tau \  F_{\mu\nu} X^{\nu}
\partial_{\tau}X^{\mu} \vert_{\sigma=0}\  +\  {1\over 2} e_2 \int d\tau\
F_{\mu\nu}
X^{\nu} \partial_{\tau}X^{\mu} \vert_{\sigma=\pi} , \cr}  \eqno(2)$$
where here the world-sheet
is a strip of width $\pi$, $\alpha'$ is the Regge slope which we will set
equal to ${1\over
2}$,  $X^{\mu}$ are $D=26$ space-time
coordinates, $F_{\mu\nu}$ is the constant field strength of a $U(1)$
embedded in the
open-string gauge group, and $e_1$, $e_2$ are the charges with respect to this
$U(1)$
which sit at the string's endpoints.
The metric both on the world sheet and in space time  is flat, and has
signature
($-+ . . . +$).
Since the external field couples to the boundary, its only effect is to
modify the
boundary conditions of the harmonic string coordinates
[\Abel]:
$$\eqalign{ \partial_{\sigma} X^{\mu} &= \  \pi  e_1 F^{\mu}_{\ \nu}
\partial_{\tau}X^{\nu}  \ \ \ \ \ (\sigma=0) , \cr
\partial_{\sigma} X^{\mu} &= \  - \pi  e_2 F^{\mu}_{\ \nu}
\partial_{\tau}X^{\nu}
\ \ \ \ \ (\sigma=\pi) . \cr}\eqno(3)$$
We will restrict our attention to a pure electric field pointing in the
$\mu=1$ direction
($F_{01}=E$) and
use light-cone coordinates: $X^{\pm} = {1\over \sqrt{2}} ( X^0 \pm X^1)$
 and ${\bf X}^{\perp}$, the latter being a $24$-component transverse
vector.
The boundary conditions then become:
$$\eqalign{ \partial_{\sigma} X^{\pm} &= \mp\beta_1 \partial_{\tau} X^{\pm}
\ \ \ \
(\sigma=0),\cr \partial_{\sigma} X^{\pm} &= \pm\beta_2 \partial_{\tau}
X^{\pm} \ \ \ \
(\sigma=\pi) ,\cr} \eqno(4a)$$
and $$ \partial_{\sigma} {\bf X}^{\perp} = 0 \ \ \ \ (\sigma=0,\pi)
,\eqno(4b)$$
where we use the notation $$\beta_{1(2)} \equiv \pi  e_{1(2)} E
\ \ . \eqno(5)$$

The transverse coordinates satisfy the usual Neumann boundary
conditions and are not affected by the presence of the field. Let us
therefore
concentrate on the light-like coordinates $X^{\pm}$.  Their mode expansion
reads
$$ X^{\pm} = x^{\pm} +  i  a_0^{\pm} \phi_0^{\pm}(\sigma,\tau) +
i
\sum_{n=1}^{\infty} \bigl[ a_n^{\pm} \phi_n^{\pm}(\sigma,\tau)
- h.c. \bigr] , \eqno(6)$$
where the orthonormal oscillation modes are [\Abel, \Bur, \Russ]
$$ \phi_n^{\pm}(\sigma,\tau) = (n\pm i\epsilon)^{-{1\over 2}}\  e^{-i(n\pm
i\epsilon)\tau} \ cos[(n\pm i\epsilon)\sigma \mp i\ arcth(\beta_1)]
\eqno(7)$$
with
$$ \epsilon = {1\over \pi} \ [arcth(\beta_1) + arcth(\beta_2)] .\eqno(8)$$
To ensure reality we must impose the hermiticity conditions
$(a_0^{\pm})^*= \pm i a_0^{\pm}$.
 Canonical quantization leads to the commutation relations:
$$ [a_n^+, (a_n^-)^*] = [a_n^-, (a_n^+)^*] = -1 ,\ \ [x^+, x^-] = {-i\pi
\over
\beta_1+\beta_2} . \eqno(9)$$
With the help of eqs. (6), (7) and (9) one can calculate
easily the Virasoro generators of
the light-like coordinates, $\{ L_n^{\ \parallel}\}$,
and check that they close an algebra with central charge
$c=2$. This is expected since the external field
does not modify the bulk properties on the world sheet. It does, however,
modify the energy of the vacuum as can be read off from the zeroth-moment
generator
$$ L_0^{\ \parallel} = -\sum_{n=1}^{\infty} (n - i\epsilon)  (a_n^+)^*a_n^-
\ - \
 \sum_{n=0}^{\infty} (n + i\epsilon)  (a_n^-)^*a_n^+ \ + \ {1\over
2}i\epsilon (1 -
i\epsilon) . \eqno(10)$$
The constant subtraction, which
 depends on our (arbitrary) choice of
 $a_0^+$ as a
destruction operator, is required in order to put the algebra
in standard form [\Abel, \Russ].
We may summarize our discussion so far, by noting that the net effect
of the electric field
is to give an
imaginary shift  $\pm i\epsilon$  to the
oscillation frequencies of the light-like coordinates,
to modify the commutation
relations of their zero modes: $x^{\pm}$, $a_0^{\pm}$, and to
change the subtraction of the vacuum. \vskip 0.3cm

We are now ready to compute the
 one-loop energy density of the vacuum,
 ${\cal F} $, whose imaginary part gives the rate
of pair creation per unit volume,
$$ w = -2 Im{\cal F} . \eqno(11)$$
Consistency forces us to consider both closed and open unoriented strings,
 and fixes the gauge group to be $SO(N)$ with $N=2^{D/2}$
 \REF\GS{M.B.Green and J.H.Schwarz, Phys.Lett. {\bf 149B} (1984) 117, and
{\bf 151B}
(1985) 21.} \REF\Weinberg{M.J.Douglas and B.Grinstein, Phys.Lett.{\bf 183B}
(1987) 52;
\hfil\break
 S.Weinberg, Phys.Lett. {\bf 187B} (1987) 278;
 \hfil\break
 N.Marcus and A.Sagnotti,  Phys.Lett.{\bf 188B} (1987) 58.}
 [\GS, \Weinberg]\footnote*{This does not of course solve the tachyon
problem of the
bosonic string.}.
 The full free energy is thus a sum of contributions of the
torus, Klein bottle, annulus and M\"obius strip,
$$ -i {\cal F} V^{(D)}   = {\cal T} + {\cal K} + {\cal A} + {\cal M}\ ,
\eqno(12)$$
where $V^{(D)}$ is the volume of space-time. Since closed-string states do
not
couple to the external field, the torus and Klein-bottle contributions are
equal to their zero-field limits: $${\cal T}={\cal T}_0 \ \ ; \ \ {\cal
K}={\cal K}_0 ,\eqno(13)$$ and will be of no interest to us here. The
contribution of the annulus for fixed values of the boundary charges reads
$$ {\cal A}(e_1,e_2) = -{1\over 2} \int_0^\infty {dt\over t}\
 Tr e^{-\pi t (L_0 - 1)}\Bigr\vert_{\sevenrm (e_1,e_2)} \ ,\eqno(14)$$
where $t$ is the modulus of the
annulus,
$$L_0 = L_0^{\perp}+L_0^{\parallel}+L_0^{ghost}\eqno(15)$$
with $L_0^{\parallel}$ given by
eq.(10), and the trace is over all string states in the $(e_1,e_2)$ charge
sector.
The full annulus amplitude is of course a sum over allowed endpoint
charges.
In order to perform this sum, note that the background electric field picks
out some
$U(1)$ direction inside the non-abelian gauge group $SO(N)$. This group is
obtained by
putting at the string endpoints ``quarks" in the ${\bf N}$ representation
of
$SO(N)$, and then performing a $Z_2$ projection to states even under the
string
reflection
 $\sigma \rightarrow \pi-\sigma$. The four terms in eq.(12) can
in fact be elegantly interpreted as corresponding to a world-sheet orbifold
construction
\REF\Sagnotti{A.Sagnotti, in Non-perturbative Quantum Field Theory, G.'t
Hooft et al
eds., Plenum 1988, p.521. } [\Sagnotti]:
the torus and annulus give the contributions of ``untwisted" and ``twisted"
sectors, in which the projection to even states is performed respectively
by the addition
of the Klein bottle and M\"obius strip. A little thought will thus convince
the reader
that the full annulus amplitude reads
$$ {\cal A} = {1\over 2} \sum_{e_1, e_2\in Q} {\cal A}(e_1, e_2) ,
\eqno(16)$$
where $Q$ is the set of charges in the decomposition of the ``quark"
representation ${\bf
N}$ under the $U(1)$ of the electric field.
Likewise the  M\"obius contribution reads
$${\cal M}\equiv {1\over 2} \sum_{e\in Q} {\cal M}(e)
= -{1\over 4} \sum_{e\in Q} \int_0^\infty {dt\over t}\
 Tr\  {\cal R} e^{-\pi t (L_0 - 1)}\Bigr\vert_{\sevenrm (e,e)}\ \ ,
\eqno(17)$$
where ${\cal R}$ is the string-reflection operator. Note that since this
operator
interchanges
string endpoints, only strings with identical endpoint charges contribute
to the
trace. Alternatively, this restriction follows from the fact that the
M\"obius strip has
a single connected boundary.
\vskip 0.3cm

We turn now to the evaluation of the traces in eqs. (14) and (17). These
factorize into the
contribution of transverse  coordinates, and light-like coordinates plus
ghosts.
Let us consider first the annulus.
Since transverse coordinates
 are not affected by the electric field, we find as usual
 $$ Tr e^{-\pi t (L_0^{\perp} - 1)} =  \int {d{\bf x}^{\perp}d{\bf
p}^{\perp}
\over (2\pi)^{D-2}}\  e^{-{\pi t \over 2}{\bf p}_{\perp}^{\ 2}}\  Z(t) \ ,
\eqno(18)$$
where
$$ Z(t) = \sum_{oriented\atop states\ S} e^{-{\pi t \over 2} M_S^{\ 2}}
\eqno(19)$$
 is the partition function (in some fixed charge sector)
 of oriented open-string states,
  weighted with the square of their mass $M_S$ .
In the critical dimension one has \break
$Z(t) = \eta({it\over 2})^{-24}$,
with $\eta(\tau)$  the Dedekind eta function.
The factorization of the
transverse contribution, as well as eqs. (18) and (19) are, however, valid
more generally
for any compactification which does not break the $SO(N)$ gauge symmetry.
Thus $D$
will   stand in the sequel for the dimension of uncompactified space time.
The
contribution of the light-like coordinates and ghosts can be obtained
easily
with the help of eqs. (9)
and (10), and reads
$$\eqalign{
Tr e^{-\pi t (L_0^{\parallel}+L_0^{ghost})}\Bigr\vert_{\sevenrm (e_1,e_2)}
= \Bigl[-{\beta_1+\beta_2\over 2\pi^2}\int & dx^+ dx^-  \Bigr]
\Bigl[{e^{-\pi t {i\epsilon\over 2}(1-i\epsilon)}
  \over 1- e^{-\pi ti\epsilon}}\Bigr] \times \cr
 & \prod_{n=1}^{\infty}{(1-e^{-\pi t n})^2 \over
  (1-e^{-\pi t(n+i\epsilon)})
(1-e^{-\pi t(n-i\epsilon)}) }\cr} \ .\eqno(20)$$
The first term on the right-hand side
follows from the anomalous commutation relation
of $x^-$ and $x^+$, if one recalls the fundamental priciple of quantum
mechanics:
 given two
conjugate variables $[{\hat x}, {\hat p}] = i{h\over 2\pi}$, the
density of quantum states
reads ${d{\hat x}d{\hat p} \over h}$.
The second term in eq.(20) takes care of the vacuum subtraction, as well as
of the $n=0$
light-like oscillator. Finally the infinite product gives the contributions
of ghosts,
and of
$n\not= 0$ light-like oscillators, which cancel precisely each other in the
absence of
an external electric field.
Putting together eqs.(14), (18) and (20), and performing explicitly all but
the
$t$-integration,  we arrive at our final expression for the annulus
amplitude:
$$ {\cal
A}(e_1,e_2) = -iV^{(D)}\ {1\over 2}\int_0^\infty {dt\over t}\  (2\pi^2
t)^{-{D\over 2}}\
Z(t)\  f_A(t,\beta_1,\beta_2) \ \ ,\eqno(21)$$
where
$$ f_A = {(\beta_1+\beta_2)t
 e^{-\pi t \epsilon^2 / 2}
\over 2sin(\pi t \epsilon/2)}\ \prod_{n=1}^{\infty}{(1-e^{-\pi t n})^2
\over
  (1-e^{-\pi t(n+i\epsilon)})
(1-e^{-\pi t(n-i\epsilon)}) } \eqno(22)$$
is a field-dependent correction factor that goes to one in the {\it
zero-field} limit
($\beta_1,\beta_2\rightarrow 0$).

A couple of remarks are here in order concerning the above annulus
amplitude.
First, the case of a pure magnetic field can be obtained easily by
 analytic continuation  $\beta_{1(2)} \rightarrow -i\beta_{1(2)}$. Second,
the
contribution of the light-like coordinates is identical to that of a
twisted unprojected
sector of an orbifold with purely imaginary twist $i\epsilon$. Finally, it
is
interesting to consider the {\it neutral-string} limit:
$\beta_1+\beta_2=\delta\rightarrow 0$ with $\beta_1$ held fixed. In this
limit
 $\epsilon\pi = {\delta \over 1-\beta_1^{\ 2}} + o(\delta^2)$, so that the
correction
factor takes the simple form
$$ f_A = (1-\beta_1^{\ 2}) +o(\delta) \ \  .\eqno(23)$$
The annulus amplitude is in this case proportional to the square of the
Born-Infeld
action,  in accordance with the result of ref. [\Abel].

The contribution of the annulus to the vacuum energy ${\cal F}$ would be
real, if the
(real) integrand in eq.(21)
had no poles on the positive $t$-axis
\footnote*{The
ultraviolet divergence at $t\rightarrow 0$ can be related in the transverse
channel to
dilaton and graviton tadpoles. In the absence of an external field
these cancel between
annulus and M\"obius strip, provided the gauge group is $SO(2^{D/2})$ [\GS,
\Weinberg].
A finite electric field, on the other hand, should
modifiy the graviton and dilaton backgrounds, so
that to cancel the $t\rightarrow 0$ divergence one needs the contributions
of the disk and projective plane.
 We can consistently ignore these background modifications here,
 since their effect on pair production is
 suppressed by an extra power of the string coupling constant. }.
This is
indeed the case for  neutral strings, as should be expected from the fact
that they do not
contribute to the pair-production rate.
For charged strings on the other hand, $f_A$
has simple poles at all  $t={2k \over \vert
\epsilon\vert}$\ ($k=1,2 . .$), and the amplitude acquires an imaginary
part.
This is given by the sum of residues at the poles times $\pi$, since the
integration contour
should pass to the right of all poles, as dictated by the proper definition
of the Feynman
propagator.
Combining eqs. (11), (12), (16), (19) and (21), and using the fact that
  $$ Res\ {f_A\over t}\Bigr\vert_{t={2k\over
\vert \epsilon\vert}}\  =\  (-)^k\  {\beta_1+\beta_2\over \pi\epsilon}
e^{-\pi k
\vert\epsilon\vert}, \eqno(24)$$
  we finally obtain the
rate for charged-string pair production,
expressed as a sum over all physical states of the (unoriented) string,
including all
possible endpoint charges \footnote\dagger{To simplify notation we here
suppress the
dependence of $\beta_1, \beta_2$ and $\epsilon$ on the state $S$.},
$$ w_{bos} =   {1\over 2(2\pi)^{D-1}}\ \sum_{states \ S}
{\beta_1+\beta_2\over \pi\epsilon }
\sum_{k=1}^{\infty} (-)^{k+1}
\Bigl({\vert\epsilon\vert\over k}\Bigr)^{D/2}
\exp{\Bigl(-{\pi k \over \vert\epsilon\vert}( M_S^{\ 2}+\epsilon^2)\Bigr)}.
\eqno(25)$$
\vskip 0.2cm

The fact that in the above equation the sum runs over all physical string
states
requires some explanation. Indeed, when $e_1\not= e_2$,
one can either
symmetrize or antisymmetrize with respect to the boundary charges, so that
oriented- and unoriented-string states can be put
in one-to-one correspondence.
The latter are therefore correctly counted by the partition function,
eq.(19).
When however $e_1=e_2=e$,
oriented-string states that are odd under reflection should be projected
out of the spectrum.
 This is of course the role of  the  M\"obius diagram,
 whose contribution to the total rate has been anticipated in eq.(25).
To be more precise, let us note first that in the sector
 $\beta_1=\beta_2=\beta$,
 light-like oscillators have the same parity under string reflection as
they do in the
absence of the field:
$$ \phi^{\pm}_n(\pi-\sigma,\tau) = (-)^n \ \phi^{\pm}_n(\sigma,\tau)
.\eqno(26)$$
  Calculating the trace in eq. (17) along the same lines as for the annulus
  we find:
$$ {\cal
M}(e) = -iV^{(D)}\ {1\over 2}\int_0^\infty {dt\over t}\  (2\pi^2
t)^{-{D\over 2}}\
\Bigl( \sum_{oriented\atop states\ S} {\cal R}_S\  e^{-{\pi t \over 2}
M_S^{\ 2}}\Bigr)
f_M(t,\beta) \eqno(27)$$
\noindent where ${\cal R}_S$ is the parity of the oriented state $S$ under
reflection, and
the correction factor reads
$$ f_M = {\beta t
 e^{-\pi t \epsilon^2/ 2}
\over sin(\pi t \epsilon/2)}\ \prod_{n=1}^{\infty}{(1-(-)^n e^{-\pi t n})^2
\over
  (1-(-)^n e^{-\pi t(n+i\epsilon)})
(1-(-)^n e^{-\pi t(n-i\epsilon)}) }\ . \eqno(28)$$
Since the poles and residues of $f_M$ on the positive $t$-axis are
identical to those of
$f_A$, the imaginary part of the M\"obius amplitude will precisely complete
the projection
of states as advertized.

The pair-production rate, eq. (25), is the main result of this paper. It
reproduces
 Schwinger's classic result, eq. (1), in the weak-field limit  $\epsilon
\simeq
 (e_1+e_2)
 E + o(E^3) \ll 1$,
 if one sets $D=4$ and recalls that a particle-antiparticle pair has
$2(2J+1)$ physical states.
The departures from the field-theory result at stronger $E$ reflect
the fact that the electromagnetic coupling of strings  is not minimal
\REF\Arg{P.C. Argyres and C.R. Nappi, Phys.Lett. {\bf 224B} (1989)
89;\hfil\break
S. Ferrara, M. Porrati and V.L. Telegdi, CERN-TH. 6432/92 and UCLA/92/TEP/7
 preprint
(March 1992).}[\Arg].
They are essentially parametrized by the non-linear function   $\epsilon$
which
goes to infinity
at   a limiting value of the field, $$E_{cr} = {1\over \pi \vert max\
e_i\vert}, \eqno(29)$$
where the total
rate for pair production diverges.
This divergence is here due to the fast-rising density of string states,
which implies that
$log Z({2k\over\vert\epsilon\vert}) \simeq {2\pi\over
k}\vert\epsilon\vert$ in this limit.
As we will however show below, in the case of the fermionic string it is
the rate of
creating any {\it given} pair of charge-conjugate states
that diverges when $\epsilon\rightarrow\infty$.
The existence of a limiting field can also be deduced classically
[\Bur, \Russ].
Heuristically, above $E_{cr}$
 the string tension can no longer hold the string together.

\vskip 0.7cm

We turn now to the case of the open superstring for which the world-sheet
action reads
$$\eqalign{
S_{sustr} = -{1\over 4\pi\alpha'} \int d\sigma & d\tau \
\bigl\{ \partial_a X^{\mu}
\partial^a X_{\mu} -
 i{\bar\psi}^{\mu}\rho^{\alpha}\partial_{\alpha}\psi_{\mu} \bigr\}
\cr  + &{1\over 2} e_1 \int d\tau \  F_{\mu\nu}
\bigr\{ X^{\nu}
\partial_{\tau}X^{\mu}
-{i\over 2}  {\bar\psi}^{\nu}\rho^{0}\psi^{\mu}
\bigr\}\bigr\vert_{\sigma=0}\cr
+\ & {1\over 2} e_2 \int d\tau\  F_{\mu\nu} \bigr\{
X^{\nu} \partial_{\tau}X^{\mu}
-{i\over 2}  {\bar\psi}^{\nu}\rho^{0}\psi^{\mu}
\bigr\}\bigr\vert_{\sigma=\pi}
. \cr}  \eqno(30)$$
Here $\psi^{\mu}$ are real two-dimensional Majorana fermions,
the critical dimension is $D=10$, and our conventions for the
Dirac matrices $\rho^{\alpha}$ are as in ref.
\REF\GSW{M.B.Green, J.H.Schwarz and E.Witten,
Superstring Theory, Cambridge University Press 1987.}[\GSW].
Both fermionic and bosonic coordinates satisfy free-wave equations, and the
boundary
conditions of the latter are given as before by eq. (3). To derive the
boundary
conditions of the fermions, one must constrain their variation as follows
[\GSW]:
$$ \delta\psi_R^{\mu} = \delta\psi_L^{\mu}\ \bigr\vert_{\sigma=0} \ \ \ ; \
\
\delta\psi_R^{\mu}= -(-)^{a} \delta\psi_L^{\mu}\ \bigr\vert_{\sigma=\pi} \
\ ,
\eqno(31)$$
where $\psi_L^{\mu}$, $\psi_R^{\mu}$ are the left- and right-moving
components of the fermion,
and $a=0$ or $1$ according to whether we are in the Neveu-Schwarz or Ramond
sector.
A straightforward calculation then gives:\footnote*{In ref. [\Russ]
 it was incorrectly claimed that the external field
does not affect the boundary conditions of the fermions. The problem can be
traced to an
error in the two-dimensional Lagrangian.}
$$ \eqalign{ \ \psi_R^{\mu} - \psi_L^{\mu}
&= \pi e_1 F^{\mu}_{\ \nu}  \Bigl( \psi_R^{\nu} + \psi_L^{\nu}\Bigr) \ \ \
\ \ \ (\sigma=0) \cr
\ \psi_R^{\mu} +(-)^{a} \psi_L^{\mu} &= -\pi e_2 F^{\mu}_{\ \nu}
\Bigl( \psi_R^{\nu} -(-)^{a} \psi_L^{\nu}\Bigr) \ \ \ \ (\sigma=\pi) . \cr}
\eqno(32)$$
We confine again our attention to an electric field in the $\mu=1$
direction, and use
light-cone coordinates $\psi^{\pm}$ and $\psi^{\perp}$, in terms of which
the boundary
conditions take the following form:
$$\eqalign{
(1\mp\beta_1) \psi_R^{\pm} &= (1\pm\beta_1) \psi_L^{\pm}\ \ \ ;
\ \psi_R^{\perp} = \psi_L^{\perp}
\ \ \ \ \ \ \ (\sigma = 0) \cr  \ & \ \cr
  (1\pm\beta_2) \psi_R^{\pm} = -(&-)^{a}(1\mp\beta_2) \psi_L^{\pm}\
  \ \ ;\ \psi_R^{\perp} =
-(-)^{a}\psi_L^{\perp} \ \ \ \ (\sigma = \pi) .\cr} \eqno(33)$$
The mode expansion of the fermionic light-like coordinates reads
$$ \psi_{R,L}^{\pm} = \sum_n d_{n}^{\pm} \chi_{(n) R,L}^{\pm}(\sigma,\tau)
\ , \eqno(34)$$
where
$$\eqalign{ \chi^{\pm}_{(n) R} & = {1\over\sqrt{2}} \exp\Bigl(-i(n \pm
i\epsilon)
(\tau-\sigma)
\pm arctanh(\beta_1)\Bigr) \cr
\chi^{\pm}_{(n) L} & = {1\over\sqrt{2}}
\exp\Bigl(-i(n \pm i\epsilon)(\tau+\sigma) \mp
arctanh(\beta_1)\Bigr) ,\cr
 } \eqno(35)$$
and the index $n$ here runs over the integers or half-integers, according
to
whether we are in the Ramond or Neveu-Schwarz sectors.
The operators $d_n^{\pm}$ obey the standard hermiticity and anticommutation
relations:
$$ (d_n^{\mu})^* = d_{-n}^{\mu}\ \ ; \ \
\{ d_n^{\mu}, d_r^{\nu}\} = \eta^{\mu\nu} \delta_{n+r} \ . \eqno(36)$$
The zero-moment Virasoro generator for the light-like fermionic coordinates
reads
$$ L_{0,ferm}^{\parallel} = -\sum_{n\in{\bf Z}+{a+1\over 2}}
 (n+i\epsilon) : d_{-n}^- d_n^+: +\  c(a) \ ,\eqno(37)$$
where
$$\eqalign{ c(0) &= -\epsilon^2/ 2  \ \ \ \ \cr
  c(1) &= {1\over 8} -{i\epsilon\over 2}(1-i\epsilon) \ \ \ \cr}
\eqno(38)$$
Notice that in the Ramond sector the field-dependent subtraction cancels
between bosons and
fermions as required by world-sheet supersymmetry, while the subtraction in
the Neveu-Schwarz
sector can be obtained easily by spectral flow. In summary,
  the effect of the electric field is to ``twist" the
fermionic light-like coordinates in the same way as the bosonic ones.

\vskip 0.3cm

We may now proceed to calculate the one-loop energy density of the
vacuum.  For fixed endpoint charges the annulus amplitude reads
$$ {\cal A}(e_1,e_2) = -{1\over 2} \int_0^\infty {dt\over t}\
\sum_{a,b = 0,1}
 C{a\atopwithdelims[] b}\
 Tr\   (-)^{bF} e^{-\pi t (L_0 - 1)}
 \Bigr\vert_{ (e_1,e_2\vert a)} \ ,\eqno(39)$$
where $L_0$ is the total Virasoro generator of bosonic and fermionic
coordinates and
ghosts, $F$ is the world-sheet fermion number, and
$ (e_1,e_2\vert a)$ is the sector of
states with boundary conditions given by eq. (33).
The coefficients in the sum over spin structures are chosen as usual:
$$C{0\atopwithdelims[] 0}=
-C{0\atopwithdelims[] 1}= -C{1\atopwithdelims[] 0} = \pm
C{1\atopwithdelims[] 1} = 1/2 \ \ ,
\eqno(40)$$
so as to enforce the GSO projection, and fix the chirality of space-time
spinors.
We will restrict ourselves to the critical uncompactified model,
since compactification is not essential for our purposes
 but would render the equations more obscure.
Following the same steps as in the case of the bosonic string we can
compute the
trace in eq. (39), and express the amplitude in the following form
$$\eqalign{ {\cal
A}(e_1,e_2) = -iV^{(D)}\ {1\over 2}\int_0^\infty {dt\over t}\  &(2\pi^2
t)^{-5}
\
\eta({it\over 2})^{-8}\  f_A(t,\beta_1,\beta_2) \times \cr
&   \sum_{a,b = 0,1}
 C{a\atopwithdelims[] b}\
 \Bigl\{  {\Theta{a\atopwithdelims[] b}(0\vert {it\over 2})
 \over \eta({it\over 2})} \Bigr\}^{4}\
 g{a\atopwithdelims[] b}(t,\epsilon)  \ . \cr}
\eqno(41)$$
The terms outside the summation sign give
the contribution of the ten bosonic coordinates,  with $f_A$ defined by eq.
(22).
The theta functions come from the trace over transverse
fermionic coordinates, which are not affected by the presence of the field:

$$ {\Theta{a\atopwithdelims[]b}(0\vert {it\over 2})\over \eta({it\over 2})}
=
   q^{{ a^2\over 8}-{1\over 24}} \prod_{n=1}^{\infty}
(1+q^{n+ {a-1\over 2}} e^{i\pi b })
\prod_{n=1}^{\infty}
(1+q^{n- {a+1\over 2}} e^{i\pi b  })  ,
\eqno(42)$$
where $q\equiv e^{-\pi t }$.
Finally $g{a\atopwithdelims[] b}$ is a  correction
factor, equal to  the contribution of light-like
fermionic coordinates and superghosts.
Using eqs. (36-38) one can easily check that for the even spin structures
($ab=0$),
$$ g{a\atopwithdelims[] b}(t,\epsilon)
= {\Theta{a-2i\epsilon\atopwithdelims[] b}(0\vert {it\over 2})
\over
\Theta{a\atopwithdelims[] b}(0\vert {it\over 2}) } \
\  .
\eqno(43)$$
This expression is not valid for the odd spin structure
($a=b=1$), for which one must first extract
appropriately the superghost zero mode. This will not
however be necessary here, because the
contribution of the odd spin structure
to the amplitude vanishes, thanks to the transverse fermionic zero
modes.

The entire field dependence of the amplitude, eq. (41), is hidden in the
correction factors
$f_A$ and $g{a\atopwithdelims[] b}$.
In the zero-field limit ($\beta_1,\beta_2,\epsilon\rightarrow 1$) one finds
$f_A$, $g{a\atopwithdelims[] b}\rightarrow 1$,
and the amplitude  reduces
  to its usual form,
$$ {\cal
A}(e_1,e_2) \rightarrow -iV^{(D)}\ {1\over 2}\int_0^\infty {dt\over t}\
(2\pi^2 t)^{-{D\over
2}}\ \Bigl[ Z_{bos}(t) - Z_{ferm}(t) \Bigr]\  \ = 0  \ \ .\eqno(44)$$
Here
$Z_{bos(ferm)}$ is defined as in eq. (19), with the sum running over
  oriented Neveu-Schwarz (Ramond) open-string states, and
$Z_{bos}=Z_{ferm}$ by virtue of
space-time supersymmetry.
For finite electric field, on the other hand,
supersymmetry is explicitly broken, so that the amplitude ${\cal
A}(e_1,e_2)$ does not
generically vanish. Now  simple inspection shows that
$g{a\atopwithdelims[] b}$ has no
poles on the positive-$t$ axis.
The only poles of the integrand are therefore those of the
function $f_A$ at $t={2k\over \vert\epsilon\vert}$, and the rate for pair
creation is
obtained by summing the residues at these poles,
precisely as in the case of the bosonic string.
The only novel input
 comes from the values of the
spin-structure-dependent correction factors at these poles,
$$g{a\atopwithdelims[] b}({2k\over\vert\epsilon\vert}, \epsilon) = (-)^{ak}
e^{\pi k
\vert\epsilon\vert} \ \ .\eqno(45)$$
Their effect on the residue is to give  an extra alternating sign to Ramond
states, and
to cancel the
effective quadratic mass shift occuring in eq. (25).
Skipping further details, let us then
 give the final expression for the pair-production rate of open
superstrings:
$$ w_{sustr} =   {1\over 2(2\pi)^{D-1}}\ \sum_{states \ S}
{\beta_1+\beta_2\over \pi\epsilon }
\sum_{k=1}^{\infty} (-)^{(k+1)(a_{S}+1)}
\Bigl({\vert\epsilon\vert\over k}\Bigr)^{D/2}
\exp{\Bigl(-{\pi k  M_S^{\ 2}\over \vert\epsilon\vert}  \Bigr)},
\eqno(46)$$
where $a_S =0$ or $1$ according to whether $S$ is in the  Neveu-Schwarz
or Ramond  sector.

Although we have only sketched its derivation in the critical dimension $D=10$,
the above expression remains valid even after compactification.
Setting $D=4$, and recalling that
    $a_S\equiv 2J$(mod $2$)  stands for the
space-time fermion number, one can recover easily Schwinger's result in the
weak-field limit.
Departures from field theory are parametrized, as in the bosonic string, by
the non-linear
function $\epsilon$, which goes to infinity at the limiting field value,
eq. (29).
Due, however, to the absence of a quadratic mass shift, the pair-production
rate of
superstrings
 diverges near $E_{cr}$ for every {\it individual} pair of charge-conjugate
states.

We conclude with two more remarks.
First, after cancelling the $t\rightarrow 0$ divergence
 with
appropriate graviton and dilaton insertions on the disk and projective
plane,
one can in principle extract
from our expressions the   induced (one-loop) ''Euler-Heisenberg"
Lagrangian of the theory.
Using the same techniques it
should, in particular,  be possible to calculate open-string threshold
effects,
as well as the index of the open-string Dirac-Ramond operator.
Finally, though carried
out in the context of fundamental strings, our calculation of pair
production
  could also apply to QCD mesons.

\vskip 0.3cm
We thank A. Sagnotti for many conversations on open strings. One of us
(C.B.) aknowledges travel support from EEC grant SC1-0394-C.

\refout

\end